\documentstyle[12pt]{article}
\begin{document}
\begin{flushright}
IC/97/165 \\
hep-ph/9711XXX
\end{flushright}
\vspace*{0.5cm}

\begin{center}
{\bf SCALAR SECTOR OF THE  3 3 1 MODEL\\
 WITH THREE HIGGS TRIPLETS}\\
\vspace*{1cm}

{\bf Hoang Ngoc Long}\\
\vspace*{0.5cm}

{\it  Institute of  Physics,
National Centre for Natural Science and 
Technology,\\
P. O. Box 429, Bo Ho, Hanoi 10000, Vietnam}\\
\vspace*{0.5cm}

and

\vspace*{0.5cm}

{\it International Centre for Theoretical Physics, 
Trieste 34100, Italy.}
\vspace*{1cm}

{\bf Abstract}\\
\end{center}

A scalar sector of the 3 3 1 model with three Higgs triplets
is considered. The mass spectrum, eigenstates and interactions
of the Higgs and the  SM gauge bosons  are derived. We show that 
one of the neutral scalars can be identified with the 
standard model Higgs boson, and in the considered potential
there is no mixing between scalars having
VEV and ones without VEV.

PACS number(s): 11.15.Ex, 12.60.Fr, 14.80.Cp\\

\newpage
\noindent
{\large\bf I. Introduction}\\[0.3cm]
\hspace*{0.5cm}The standard model (SM) of the electroweak interactions 
has been considered an extremely successful theory from the 
phenomenological point of view. However, this theory contains
a large number of free parameters as well as a large number
of unanswered questions. These drawbacks of 
the SM have led to a strong belief 
that the model is still incomplete. One of the basic elements of the SM  
has not been tested, that is an observation of the Higgs scalar. 
At present, hunting Higgs
boson and looking for a new physics beyond the SM are central tasks of 
particle physics. Hopefully, these tasks can be 
done at LEP2, Tevatron, and others.

In recent years, the study of scalar sector has become one of the
booming subjects in particle physics. This study has been 
carried out within the SM 
framework as well as some extensions of the SM. 
One of these extensions is
the $\mbox{SU}(3)_C\otimes \mbox{SU}(3)_L \otimes \mbox{U}(1)_N$ 
(3 3 1) models~\cite{ppf,rf}. 
These models have the following intriguing features:
Firstly, the models are anomaly free only if the number
of families N is a multiple of three. If further
one adds the condition of QCD asymptotic freedom, 
which is valid only if the number of families 
of quarks is to be less than five, it follows that N 
is equal to 3. The second characteristic of these models
is that one family of quarks is treated differently
from the other two. This could lead to a natural explanation
for the unbalancing heavy top quarks, deviations of $A_b$
from the SM prediction etc. In addition, the models
predict no very high new mass scales -- a few TeV~\cite{dng}.

Three Higgs triplets
of the minimal version~\cite{ppf} have  firstly been analysed,
then the sextet has been added~\cite{rf} in the further consideration. 
A further study of this model in Higgs sector was done 
recently by Tonasse~\cite{ton}.
In the minimal 3 3 1 model it is necessary to have three 
triplets (in which there are two doubly charged and  three neutral
scalar states) and sextet in order for leptons to gain masses
(for more details, see~\cite{rf}). However, in an 
alternative 3 3 1 model which
has {\it neutral} leptons such as neutrinos [5 -- 8]  
in triplet, the Higgs sector has a quite different
content, i.e. more neutral scalars (five) and no doubly charged Higgs
states. Consequently, the Higgs potential constructed on these
states, will have own features.

In the present paper we  study the scalar sector of the model
with {\it real}  three Higgs triplets.
This paper is organized as follows. In Sec. II, the main 
notations and the potential are introduced. Using obtained
constraint equations, the mass spectrum and eigenstates
in the potential are derived. In order to compare with
the SM Higgs structure,
Sec. III is devoted to couplings
of the obtained Higgs  with the SM gauge bosons such as
 $Z,  W^\pm$ and the photon $\gamma$. We show that under some
circumstances we can get  complete analogy with 
coupling constants in the SM. Finally, our conclusions 
are summarized in the last section.\\[0.3cm]
{\large\bf II. The potential and mass spectrum}\\[0.3cm]
\hspace*{0.5cm}The details of the 3 3 1 models with neutral
lepton in the bottom of  the leptonic triplet were pointed
out in  Refs. [5 -- 8]. Here we remind only symmetry breaking in
these  models
\begin{eqnarray}
&\mbox{SU}(3)_{C}&\hspace*{-0.2cm}\otimes \ \mbox{SU}(3)_{L}\otimes 
\mbox{U}(1)_{N}\nonumber \\
&\downarrow      &\hspace*{-0.8cm}\langle \chi \rangle   \nonumber \\
&\mbox{SU}(3)_{C}&\hspace*{-0.2cm}\otimes \ \mbox{SU}(2)_{L}\otimes 
\mbox{U}(1)_{Y}\nonumber \\
&\downarrow      &\hspace*{-0.8cm}\langle \rho \rangle, \langle 
\eta \rangle   \\
&\mbox{SU}(3)_{C}&\hspace*{-0.2cm}\otimes \ \mbox{U}(1)_{Q},
\nonumber
\end{eqnarray}
where
\begin{equation}
\chi = \left( \begin{array}{c}
                \chi'^{o}\\ \chi^-\\ \chi^o\\ 
                \end{array}  \right) \sim (1, 3, -\frac{1}{3}),
\eta = \left( \begin{array}{c}
                \eta^o\\ \eta^-\\ \eta'^{o}\\ 
                \end{array}  \right) \sim (1, 3, -\frac{1}{3}),\
\rho = \left( \begin{array}{c}
                \rho^+\\ \rho^o\\ \rho'^{+}\\ 
                \end{array}  \right) \sim (1, 3, \frac{2}{3}).
\label{hig}
\end{equation}

From Eq.~(\ref{hig}) we see that in our case we have two extra neutral
states, therefore in the neutral scalar sector we have 5 x 5 mass
matrix instead of 3 x 3 one as in~\cite{ton}. Note that $\chi$ and
$\eta$ transform the same way, therefore the most general
Higgs potential (i.e. the one including all terms consistent with
the gauge invariance and renormalizability)
is very complicated. However, as mentioned in~\cite{flt}
under assumption of  
the discrete symmetry  $\chi \rightarrow - \chi$,
the most general potential can then be written in the 
following form:
\begin{eqnarray}
V(\eta,\rho,\chi)&=&\mu^2_1 \eta^+ \eta +
 \mu^2_2 \rho^+ \rho +  \mu^2_3 \chi^+ \chi +
\lambda_1 (\eta^+ \eta)^2 + \lambda_2 (\rho^+ \rho)^2 +
\lambda_3 (\chi^+ \chi)^2 \nonumber \\
& + & (\eta^+ \eta) [ \lambda_4 (\rho^+ \rho) +
\lambda_5 (\chi^+ \chi)] + \lambda_6 (\rho^+ \rho)(\chi^+ \chi) +
\lambda_7 (\rho^+ \eta)(\eta^+ \rho) \nonumber\\
& + & \lambda_8 (\chi^+ \eta)(\eta^+ \chi) +
\lambda_9 (\rho^+ \chi)(\chi^+ \rho) +
 \lambda_{10} (\chi^+ \eta + \eta^+ \chi)^2.
\label{pot}
\end{eqnarray}

The last term in ~(\ref{pot}) reflects the fact that
$\chi$ and $\eta$ transform similarly and this term will give 
specific interactions.
For convenience in reading, we rewrite the  
expansion of the scalar fields which acquire a VEV:
\begin{equation}
\eta^o = v + H^o_\eta + i A^o_\eta ; \ 
\rho^o = u + H^o_\rho + i A^o_\rho;\
\chi^o = w + H^o_\chi + i A^o_\chi.
\label{exp1}
\end{equation}

For the sake of simplicity, here we assume that 
vacuum expectation values (VEVs) are real. This means that
the CP violation through the scalar exchange is not considered
in this work. For the  prime neutral fields which do 
not have VEV, we get analogously:
\begin{equation}
\eta'^{o} = H'^{o}_\eta + i A'^{o}_\eta ; \ 
\chi'^{o} = H'^{o}_\chi + i A'^{o}_\chi.  
\label{exp2}
\end{equation}
In literature, a real part $H$ is called  
CP-even scalar or {\it scalar},
and an imaginary one  $A$  -  CP-odd scalar or {\it pseudoscalar} field.
In this paper we call them  scalar and pseudoscalar, respectively.

The VEV $\langle \chi \rangle$  will generate masses for
exotic 2/3 and --1/3 quarks and new heavy gauge bosons $Z', X, Y$,
while VEV $ \langle \rho \rangle,  
\langle \eta \rangle$ will generate masses for ordinary fermions
and the SM gauge bosons $Z,\ W^\pm$. 
To keep  the model consistent with low-energy phenomenology,
the VEV  $\langle \chi \rangle$ must be large enough.
In this paper we will use the following approximation:
\begin{equation}
w \gg v, u.
\label{lecod}
\end{equation}

Requiring that in the shifted potential $V$, the linear terms in fields
must be absent, we get  in the tree level approximation,
 the following constraint equations:
\begin{eqnarray}
\mu^2_1 + 2 \lambda_1 v^2 + \lambda_4 u^2 + \lambda_5 w^2
 & = & 0 \nonumber\\
\mu^2_2 + 2 \lambda_2 u^2 + \lambda_4 v^2 + \lambda_6 w^2
 & = & 0 
\label{cont}\\
\mu^2_3 + 2 \lambda_3  w^2 + \lambda_5 v^2 + \lambda_6 u^2
 & = & 0\nonumber .
\end{eqnarray}

Substituting  Eqs.~(\ref{hig}),~(\ref{exp1}) 
and~(\ref{exp2}) into Eq.~(\ref{pot}) and diagonalizing,
 we will get a mass spectrum of 
Higgs bosons with mixings.\\[0.3cm]
\hspace*{0.5cm}a. {\it Spectrum in  neutral scalar sector}\\[0.3cm]  
\hspace*{0.5cm}In the $H^o_\eta, H^o_\rho, H^o_\chi, 
H'^{o}_\eta, H'^{o}_\chi$ basis
the square mass matrix, after imposing of the constraints 
~(\ref{cont}), has 
a quasi-diagonal form as follows:
\begin{equation}
M^2_H = 
\left( \begin{array}{cc}
M^2_{3H}& 0\\
0 & M^2_{2H} \end{array} \right),
\end{equation}
where 
\begin{equation}
M^2_{3H} = 
2 \left( \begin{array}{ccc} 
 2 \lambda_1 v^2  & \lambda_4 v u   &
  \lambda_5 v w   \\
  \lambda_4 v u   &  2 \lambda_2 u^2 &
  \lambda_6 u w \\
  \lambda_5 v w  &  \lambda_6 u w  &
 2 \lambda_3 w^2  \end{array} \right),
\label{mat1}
\end{equation}
and
\begin{equation}
M^2_{2H} = 
( \lambda_8 + 4 \lambda_{10} )\Biggl( \begin{array}{cc} 
  w^2    & v w \\
 v w & v^2 \end{array} \Biggr).
\label{mat2}
\end{equation}

The above mass matrix shows that the prime fields mix themselves
but do not mix  with others. 

Now we consider 3 x 3 mass matrix $M^2_{3H}$ of 
 $H^o_\eta, H^o_\rho, H^o_\chi$ mixing. In its exact form 
it is impossible to find the acceptable solution of  
the characteristic equation. However, keeping only terms of the second
order of $w$ we get immediately two massless states and one physical
field with mass $ - 4 \lambda_3 w^2$. 

In order to improve the solution we add to our consideration 
the linear terms  in $w$.
We obtained then one massless field $H_1$ with an eigenstate
\begin{equation}
H_1  \approx \frac{1}{(\lambda_5^2 v^2 +
\lambda_6^2 u^2)^{1/2}}
\left( \lambda_6 u H^o_\eta -  \lambda_5 v H^o_\rho  \right).
\label{root1}
\end{equation}

Two remaining states are $ H_2$ with mass approximatively equal to
zero and one massive physical state  $H_3$ with mass:
\begin{equation}
  m^2_{H_3} \approx - 4 \lambda_3 w^2 .
\label{mass1}
\end{equation}

The conditions for orthogonality and normality allow us to get
expressions for $H_2$ and $H_3$:
\begin{equation}
H_2  \approx \frac{1}{(\lambda_5^2 v^2 +
\lambda_6^2 u^2)^{1/2}}
\left( \lambda_5 v H^o_\eta +  \lambda_6 u H^o_\rho  \right),
\label{root2}
\end{equation}
\begin{equation}
H_3 \approx  H^o_\chi .
\label{root3}
\end{equation}

In order to gain mass for $H_1$, we
 solve the characteristic  equation with the exact
3 x 3 mass matrix $M^2_{3H}$, and the  $H_1$ associated with, namely:
\begin{equation}
\left( M^2_{3H} - I m^2_{H_1} \right) H_1 = 0.
\label{eigvq1}
\end{equation}

Solving a system of three equations ~(\ref{eigvq1})
 we get the mass for $H_1$
\begin{equation}
m^2_{H_1} \approx  \frac{v^2}{\lambda_6}(
2 \lambda_1 \lambda_6 -  \lambda_4 \lambda_5) \approx
 \frac{u^2}{\lambda_5}(
2 \lambda_2 \lambda_5 -  \lambda_4 \lambda_6).
\label{rt1}
\end{equation}

The above equation shows that mass of $H_1$ depends only on $v$ or
$u$ separately. Similarly, for $H_2$ we get
\begin{equation}
m^2_{H_2} \approx  2 \lambda_1 v^2 +  \frac{\lambda_4 
\lambda_6 u^2}{\lambda_5} \approx
 2 \lambda_2 u^2 +  \frac{\lambda_4 
\lambda_5 v^2}{\lambda_6}.
\label{rt2}
\end{equation}

Eqs.~(\ref{rt1}) and~(\ref{rt2}) also give us relations
among coupling constants and VEVs.
  
Diagonalization of 2 x 2 mass matrix $M^2_{2H}$ 
gives us one Goldstone $G'_1$ and one physical massive field
$H'_4$  with mass:
\begin{equation}
m^2_{H'_4} =  -( \lambda_8 + 4 \lambda_{10}) (v^2 + w^2),
\label{mass2}
\end{equation}
and mixing
\begin{equation}
\left(\begin{array}{c}
H'^{o}_\eta\\
H'^{o}_\chi\\
\end{array}\right) = \frac{1}{(v^2 + w^2)^{1/2}}
\left( \begin{array}{cc} 
- v & w\\
w & v\end{array} \right)
\left(\begin{array}{c}
G'_1 \\
H'_4 \end{array} \right).
\end{equation}
 
b. {\it Spectrum in neutral pseudoscalar sector}\\[0.3cm]
\hspace*{0.5cm}Now we consider  the pseudoscalar  sector.
Keeping one's mind on the constraint equations~(\ref{cont})
we have three Goldstone  bosons which can be identified as 
follows:
$ G_2 \equiv A^o_\eta,\ G_3 \equiv  A^o_\rho,\
G_4 \equiv  A^o_\chi$
and in the $A'^o_\eta, A'^o_\chi$ basis 
\begin{equation}
M^2_{2A} = 
( \lambda_8 + 4 \lambda_{10} )\Biggl( \begin{array}{cc} 
  w^2    & v w \\
 v w & v^2 \end{array} \Biggr).
\label{mat3}
\end{equation}
We easily  get one Goldstone  $G'_5$ and one massive 
pseudoscalar boson $A_1$ with mass
\begin{equation}
m^2_{A_1} =  -( \lambda_8 + 4 \lambda_{10}) (v^2 + w^2),
\end{equation}
and mixing
\begin{equation}
\left(\begin{array}{c}
H'^{o}_\eta\\
H'^{o}_\chi\\
\end{array}\right) = \frac{1}{(v^2 + w^2)^{1/2}}
\left( \begin{array}{cc} 
- v & w\\
w & v\end{array} \right)
\left(\begin{array}{c}
G'_5 \\
A_1 \end{array} \right).
\end{equation}

c. {\it Spectrum in charged scalar sector}\\[0.3cm]
\hspace*{0.5cm}In the charged sector we have two Goldstone
bosons and two physical massive fields with mixings:
\begin{equation}
\left(\begin{array}{c}
\eta^+\\
\rho^+\\
\end{array}\right) = \frac{1}{(v^2 + u^2)^{1/2}}
\left( \begin{array}{cc} 
- v & u\\
u & v\end{array} \right)
\left(\begin{array}{c}
G^+_6 \\
H^+_5 \end{array} \right),
\end{equation}
\begin{equation}
\left(\begin{array}{c}
\rho'^+\\
\chi^+\\
\end{array}\right) = \frac{1}{(u^2 + w^2)^{1/2}}
\left( \begin{array}{cc} 
- u & w\\
w & u\end{array} \right)
\left(\begin{array}{c}
G^+_7 \\
H^+_6 \end{array} \right).
\label{mix4}
\end{equation}
The masses of $H^+_5$ and $H^+_6$ are given, respectively:
\begin{equation}
m^2_{H^+_5} = - \lambda_7( v^2 + u^2),\hspace*{0.5cm}
m^2_{H^+_6} = -  \lambda_9( v^2 + w^2).
\label{mass3}
\end{equation}

We emphasize that masses of $H^+_5$ and $H_1,\ H_2$ depend
only on VEVs of light Higgs fields: $u$ and $v$.

In the low-energy phenomenology -- the limit ~(\ref{lecod}), 
the mixing in ~(\ref{mix4})
becomes small, then we have: $\rho'^+ \sim H_6^+$ and
$\chi^+ \sim G^+_7$.

Requiring that square mass of the physical fields is 
positive (otherwise, they are Goldstone ones) and combining 
Eqs.~(\ref{mass1}), ~(\ref{mass2}) and ~(\ref{mass3})
we get the following relations among parameters of the potential:
\begin{equation}
\lambda_3  \stackrel{<}{\sim}  0,\hspace*{0.2cm}
\lambda_7  \stackrel{<}{\sim}  0,\hspace*{0.2cm}
\lambda_9  \stackrel{<}{\sim}  0,\hspace*{0.2cm}
\lambda_8 + \lambda_{10}  \stackrel{<}{\sim}  0. 
\end{equation}

Now we briefly list  the  particle content in our Higgs sector: in
the considered models we have
four neutral scalars, one neutral pseudoscalar, two charged scalars
and seven Goldstone bosons.

Note that the mass spectrum and eigenstates in this sector are exact.
\\[0.3cm]
{\large\bf III. Higgs -- SM gauge boson couplings}\\[0.3cm]
\hspace*{0.5cm}In order to identify the considered above Higgs bosons
with those in the SM, in this section we present the couplings
of two kinds of  mentioned particles in the
model with right-handed neutrinos~\cite{flt,hnl}. For the multi-Higgs
case,
couplings of the Higgs bosons with the fermions (Yukawa couplings)
are not equally  defined. For example, to avoid unsuppressed lepton flavour
violating processes~\cite{gw} one uses the discrete symmetry,
while other authors suggested alternative ways for heavier
quarks (for details, see~\cite{csa}).

Interactions among the gauge bosons and the Higgs ones arise from the
following pieces:
\begin{equation}
 (D^\mu {\cal H})^+(D_\mu {\cal H}), \  \cal{H} = \eta,\ \rho \ \chi,
\label{lag}
\end{equation}
in which the covariant derivatives are defined~\cite{deriv}:
\begin{equation}
D_\mu  = \partial_\mu  + i g\sum^8_{a=1} W^a_\mu .\frac{\lambda_a}{2}
+i g_N\frac{\lambda^9}{2} N B_\mu .
\label{dve}
\end{equation}
Substituting  Eq.~(\ref{hig}) into ~(\ref{dve}) then ~(\ref{lag})
we get the following trilinear couplings:
\begin{eqnarray}
g(W W H_1) & = & \frac{g^2 u v (\lambda_6 - \lambda_5)}{
(\lambda^2_5 v^2 + \lambda^2_6 u^2)^{1/2}},
\label{cu1}\\
g(W W H_2) & = & \frac{g^2  (\lambda_5 v^2 + 
\lambda_6 u^2)}{(\lambda^2_5 v^2 + \lambda^2_6 u^2)^{1/2}},
\label{cu2}\\
g(Z Z H_1) & = & \frac{g^2 u v (\lambda_6 - \lambda_5)}{
2 c_W^2(\lambda^2_5 v^2 + \lambda^2_6 u^2)^{1/2}},
\label{cu3}\\
g(Z Z H_2) & = & \frac{g^2  (\lambda_5 v^2 + 
\lambda_6 u^2)}{ 2 c_W^2 (\lambda^2_5 v^2 + \lambda^2_6 u^2)^{1/2}},
\label{cu4}
\end{eqnarray}
where $c_W \equiv \cos \theta_W$.

From Eqs.~(\ref{cu1}, \ref{cu2}, \ref{cu3}) and~(\ref{cu4}) 
we see that if 
\begin{equation}
\lambda_5 = \lambda_6,
\label{cod2}
\end{equation}
 the interactions among  $H_1$ and $Z, W$
vanish, while the interactions with $H_2$  are found to be:
\begin{equation}
g(W W H_2) = \sqrt{2} g \ m_W , \hspace*{0.2cm} 
g(Z Z H_2) = \sqrt{2} g \ m_Z / c_W,
\end{equation}
where $ m^2_W = \frac{g^2 (v^2 + u^2)}{2}, \  m_Z =
\frac{ m_W}{c_W}$.

Looking at Eq.~(\ref{pot}) we see that the condition ~(\ref{cod2})
means that quartic interactions among light Higgs states
$\eta, \rho$ and heavy one $\chi$ are the same.
Therefore this assumption is not badly based.

The quartic couplings are determined to be:
\begin{eqnarray}
g(W W H_1 H_1)  & =  & g(W W H_2 H_2) = g(W W G_2 G_2) =
g(W W G_3 G_3) = \frac{g^2}{2},\nonumber\\
g(Z Z H_1 H_1)  & =  & g(Z Z H_2 H_2) = g(Z Z G_2 G_2) =
g(Z Z G_3 G_3) = \frac{g^2}{4 c_W^2}. \nonumber
\end{eqnarray}
For charged Higgs bosons we have
\begin{eqnarray}
g(\gamma \gamma  G^+_5 G^-_5)  & =  & g(\gamma \gamma H^+_5 H^-_5) = 
e^2,\nonumber\\
g(Z Z G^+_5 G^-_5)  & =  & g(Z Z H^+_5 H^-_5) =
\frac{g^2}{4 c_W^2}( 1 - 2 s_W^2). \nonumber
\end{eqnarray}

Keeping one's mind that the VEVs and Higgs fields in this paper
are redefined (see for example,~\cite{cl}) we recognize 
 complete analogy with the SM interactions.
 
Summarizing, from couplings of the SM gauge bosons with
Higgs bosons in this model we can conclude that 
the $H_2$ can be identified with the SM  neutral Higgs boson,
while charged ones $G_5^+$ -- with charged Goldstone
boson. Another neutral Higgs boson is very light
and interacts weakly with light (i.e. the SM) gauge bosons.\\[0.3cm]
{\large\bf IV. Summary}\\[0.3cm]
\hspace*{0.5cm}In this paper we have considered the mass spectrum,
eigenstates of the potential specialized for the 3 3 1 models
with three Higgs triplets. It is shown that in the considered
model there are two light neutral Higgs bosons and one
of them can be identified with the SM Higgs.
In the some circumstance the coupling becomes identical
to that in the SM.
Other charged Higgs bosons such as scalar $H_5^+$ and pseudoscalar
$G^+_5$ interact to the SM gauge bosons including the photons $\gamma$,
with the same coupling constants.

We emphasize again that the analysis here 
(excluding the neutral scalar sector)  are exact.

The light neutral Higgs bosons play an important role in the testing
of new physics beyond the SM~\cite{bar}.

In this letter we have  analysed the
scalar sector in the 3 3 1 model with three Higgs triplets,
especially in the two parts scalar  and pseudoscalar sector.
An application in new physics beyond the SM
will be a subject of further study.

The most general potential for this kind of  3 3 1 models
is very complicated and it is impossible to get an acceptable
solution. However, with the help of
the discrete symmetry, we obtained rather simple solutions,
because  there are no mixings between scalars
having VEV and ones without VEV.

In concluding, the considered here 3 3 1 model with
three Higgs triplets has a rather  simple Higgs
structure. In addition, the data from neutrino neutral current
scatterings gives a lower bound for mass of the
heaviest new neutral gauge boson $Z'$ of few
hundreds GeV (and hence the VEV $w$)~\cite{hnl}.
With no very high new mass scales the 3 3 1 models
are interesting, and they can be confirmed or ruled
out in the near future.
   
The author thanks Dr. R. Foot for useful remarks, 
he would also like to thank
Professor S. Randjbar-Daemi and International Centre for 
Theoretical Physics, Trieste for support and hospitality.
This work was supported in part by  National Foundation
for Basic  Science Contract No: KT - 04.1.1.\\[0.3cm]


\begin{thebibliography}{9}
\bibitem{ppf} F. Pisano and V. Pleitez, Phys. Rev.  D 46, (1992) 410;\\
P. H. Frampton, Phys. Rev. Lett.  69, (1992) 2889.
\bibitem{rf} R. Foot, 
O. F. Hernandez, F. Pisano, and V. Pleitez, Phys. Rev. D 47, (1993) 4158.
\bibitem{dng}D. Ng, Phys. Rev.  D 49, (1994) 4805.
\bibitem{ton}M. D. Tonasse, Phys. Lett. B 381, (1996) 191.
\bibitem{svs}M. Singer, J. W. F. Valle, and J. Schechter, Phys. Rev.
D 22, (1980) 738.
\bibitem{flt} R. Foot, H. N. Long, and Tuan A. Tran, 
 Phys. Rev. D 50, (1994) R34.
\bibitem{hnl}H. N. Long, Phys. Rev. D 54, (1996) 4691.
\bibitem{mpp}J. C. Montero, F. Pisano, and V. Pleitez,
Phys. Rev. D 47, (1993) 2918. 
\bibitem{gw}S. Glashow and S. Weinberg, Phys. Rev.
D 15, (1977) 1958.
\bibitem{csa}T. P. Cheng and M. Sher, Phys. Rev. D 35,
(1987) 3484; D 44, (1991) 1461;\\
A. Antaramian, L. J. Hall, and A. Raisin,
Phys. Rev. Lett. 69, (1992) 1871;\\
 M. Luke and M. J.
Savage, Phys. Lett. B 307, (1993) 387.
\bibitem{deriv}J. Liu and D. Ng, Phys. Rev. D 50, (1994) 548;\\
H. N. Long, Phys. Rev. D 53, (1996) 437. 
\bibitem{cl}T. P. Cheng and L. F. Li, {\it Gauge field theory
of elementary particle physics},\ (Oxford University, New York,
1984), pp 510 -- 512;\\
P. D. B. Collins, A. D. Martin, and E. J. Squires, {\it 
Particle physics and cosmology},\ (John Willey \& sons,
New York, 1989) pp. 85 -- 97;\\
J. F. Gunion, H. E. Haber, G. Kane, and S. Dawson,
{\it The Higgs hunter's guide}, \ (Addison-Wesley
Publishing Company, Redwood City, CA, 1990) pp. 17 -- 19.
\bibitem{bar}S. Bar-Shalom, G. Eilam, A. Soni, and J. Wudka,
Phys. Rev. Lett. 79, (1997) 1217; D. Atwood, L. Reina,
and A. Soni, Phys. Rev. D 55, (1997) 3156.
\end{thebibliography}
\end{document}